\documentclass[grl]{agutex}

\usepackage{lineno}
\usepackage[dvips]{graphicx}
\setkeys{Gin}{draft=false}
\authorrunninghead{WATKINS AND CHO}

\titlerunninghead{JUPITER'S EQUATORIAL ZONAL WIND}

\authoraddr{J. Y-K. Cho,
School of Physics and Astronomy,
Queen Mary University of London,
Mile End Road, London E1 4NS, UK.
(J.Cho@qmul.ac.uk)}

\authoraddr{C. Watkins, 
School of Physics and Astronomy,
Queen Mary University of London,
Mile End Road, London E1 4NS, UK.
(C.Watkins@qmul.ac.uk)}

\begin{document}

%% ------------------------------------------------------------------------ %%
%
%  TITLE
%
%% ------------------------------------------------------------------------ %%

\title{The vertical structure of Jupiter's equatorial zonal wind above
  the cloud deck, derived using mesoscale gravity waves}
%
% e.g., \title{Terrestrial ring current:
% Origin, formation, and decay $\alpha\beta\Gamma\Delta$}
%

%% ------------------------------------------------------------------------ %%
%
%  AUTHORS AND AFFILIATIONS
%
%% ------------------------------------------------------------------------ %%

%Use \author{\altaffilmark{}} and \altaffiltext{}

% \altaffilmark will produce footnote;
% matching \altaffiltext will appear at bottom of page.

\authors{C. Watkins\altaffilmark{1}
     and J. Y-K. Cho\altaffilmark{1}}

   \altaffiltext{1}{School of Physics and Astronomy, Queen Mary
     University of London, London, UK.}

% \authors{R. C. Bales,\altaffilmark{1}
% E. Mosley-Thompson,\altaffilmark{2} R. Williams,\altaffilmark{3}
% J. R. McConnell\altaffilmark{4}, and Francesco Visconti\altaffilmark{5}}

%\altaffiltext{1}{Department of Hydrology and Water Resources,
%University of Arizona, Tucson, Arizona, USA.}

%\altaffiltext{2}{Department of Geography, Ohio State University,
%Columbus, Ohio, USA.}

%\altaffiltext{3}{Department of Space Sciences, University of
%Michigan, Ann Arbor, Michigan, USA.}

%\altaffiltext{4}{Division of Hydrologic Sciences, Desert Research
%Institute, Reno, Nevada, USA.}

%\altaffiltext{5}{Dipartimento di Idraulica, Trasporti ed
%Infrastrutture Civili, Politecnico di Torino, Turin, Italy.}

%% ------------------------------------------------------------------------ %%
%
%  ABSTRACT
%
%% ------------------------------------------------------------------------ %%

% >> Do NOT include any \begin...\end commands within
% >> the body of the abstract.

\begin{abstract}
  Data from the Galileo Probe, collected during its descent into
  Jupiter's atmosphere, is used to obtain a vertical profile of the
  zonal wind from $\mathbf{\sim\! 0.5}$~bar (upper troposphere) to
  $\mathbf{\sim\! 0.1\, \mu\mbox{bar}}$ (lower thermosphere) at the
  probe entry site.  This is accomplished by constructing a map of
  gravity wave Lomb-Scargle periodograms as a function of altitude.
  The profile obtained from the map indicates that the wind speed
  above the visible cloud deck increases with height to
  $\mathbf{\sim\!150}$~m\,s$\mathbf{^{-1}}$ and then levels off at
  this value over a broad altitude range.  The location of the
  turbopause, as a region of wide wave spectrum, is also identified
  from the map.  In addition, a cross-equatorial oscillation of a jet,
  which has previously been linked to the quasi-quadrennial
  oscillation in the stratosphere, is suggested by the profile.
\end{abstract}

%% ------------------------------------------------------------------------ %%
%
%  BEGIN ARTICLE
%
%% ------------------------------------------------------------------------ %%

% The body of the article must start with a \begin{article} command
%
% \end{article} must follow the references section, before the figures
%  and tables.

\begin{article}

%% ------------------------------------------------------------------------ %%
%
%  TEXT
%
%% ------------------------------------------------------------------------ %%

\section{Introduction}

The horizontal structure of Jupiter's zonal (east-west) winds at the
level of the visible cloud deck on Jupiter have been observed for many
decades
\citep[e.g.,][]{Limaye1986,Ingersoll2004,Garcia2001,Vasavada2005,Li2006}.
In contrast, the vertical structure of zonal winds is not well known
-- especially away from the cloud deck and in the equatorial region.
This is because of the lack of usable tracers and the breakdown of the
thermal wind relation near the equator, preventing latitudinal
temperature measurements to be related to the vertical wind shear.
  
In December 1995 the Galileo probe entered Jupiter's upper atmosphere
at 6$^{\circ}\!$~N, above a $5 \mu$m hot spot in the north equatorial
belt.  During the descent the Atmospheric Structure Instrument on
board collected information about the density, pressure, and
temperature of the atmosphere \citep{Seiff1998}.  Analysis of the
temperature profile in the thermospheric and stratospheric regions
identified discernible perturbations which have been interpreted as
manifestations of internal, or vertically propagating, gravity waves
\citep{Young1997,Young2005,Matcheva1999}.  Vertical temperature
profiles for Jupiter's atmosphere have also been obtained from
occultation studies using stars and other spacecraft.  A number of
these profiles contain oscillations that have been characterised as
manifestations of internal gravity waves as well
\citep{French1974,Lindal1992,Hubbard1995,Raynaud2003,Raynaud2004}.

Gravity waves are oscillations of fluid parcels about their altitudes
of neutral buoyancy \citep{Gossard1975}.  The waves are a common
feature of stably stratified atmospheres. They have been captured in
images of Jupiter's clouds, typically near the planet's equator
\citep{Flasar1986,Arregi2009,Reuter2007}.  Fig.~1 shows an example.
Internal gravity waves (hereafter simply gravity waves) grow in
amplitude, due to the fall in background atmospheric density.  The
dynamics of such waves is described by the Taylor--Goldstein equation
\citep{Taylor1931,Goldstein1931},
\begin{linenomath*}
\begin{equation}
%\begin{eqnarray*} 
\frac{{\rm d}^2 w}{{\rm d}z^2} + m^2(z)\, w\ =\ F(z)\, .
%\end{eqnarray*}
\end{equation}
\end{linenomath*}
Here $w$ is the perturbation in the vertical wind, which is adjusted
for the amplitude growth and also assumed to be oscillatory
(wave-like) in the horizontal direction and time; $F$ is a function
that represents the source and dissipation of waves; and, $m$ is the
vertical wavenumber, which depends on the physical properties of the
medium in which the waves propagate.  Specifically,
\begin{linenomath*}
\begin{eqnarray}
  m(z)\  = \hspace*{6.35cm} \nonumber \\
  \left[\,\frac{N^2}{I^2} - \frac{1}{I}\,\frac{{\rm d}^2 I}{{\rm
        d}z^2} - \frac{1}{HI}\,\frac{{\rm d} I}{{\rm d}z} -
    \frac{1}{4H^{2}}\left(1 - 2\frac{{\rm d}H}{{\rm d}z}\right) -
    k^2\,\right]^{1/2}\hspace*{-.2cm} , 
  \end{eqnarray} 
\end{linenomath*}
where $N$ is the Brunt--V\"ais\"al\"a (buoyancy) frequency; $I$ is the
intrinsic phase speed, $c - u_0$, where $c$ is the horizontal phase
speed and $u_0$ is the zonal wind; $H$ is the density scale height;
and, $k$ is the horizontal wavenumber.  In general, all of the
variables depend on altitude $z$.  Crucially, the zonal wind profile
$u_0 (z)$ can be obtained by solving for $I = I(H,k,N,m)$ in
equation~(2).  In this letter we present a profile from this inversion
that span much larger range of altitude above the 1~bar level (range
of $\sim\!  500$~km) than in past analyses.

\section{Method}

For the inversion we compute $H$ from the collected density profile.
We also use $k = 2\pi/300$~km$^{-1}$ from the average value obtained
in analysis of Voyager images \citep{Flasar1986}.  Other analyses have
reported slightly different values (e.g., average of
$2\pi/165$~km$^{-1}$ by \citet{Arregi2009}); however, as long as $k^2
\ll m^2$ (as is the case here), the value of $k$ does not
significantly alter the result of the inversion.  $N$ and $m$ are
obtained from the potential temperature, $\theta = T(p_{\rm
  ref}/p)^\kappa$, related to specific entropy of the atmosphere; here
$T$ is the temperature, $p$ is the pressure, $p_{\rm ref} = 1$~bar is
a constant reference pressure and $\kappa = R/c_p$ with $R$ the
specific gas constant and $c_p$ the specific heat at constant
pressure.  To obtain $N$ and $m$, we decompose $\theta$ into a mean
background $\bar{\theta}$ (extracted using a spatially-moving fitting
window) and a small perturbation $\Delta \theta$ about the mean
(presumed caused by mesoscale gravity waves).  Such decomposition of
temperature is standard in gravity wave studies
\citep{Lindzen1990,Nappo2002,Young1997}, and is reasonable here given
that the spatial scales of the background and waves are well
separated.  Further, the horizontal distance travelled by the probe in
the region studied, 3300~km, is small compared to Jupiter's
circumference and is completely within the hot-spot entered by probe
\citep{Orton1998}; hence, zonal variation in the background over this
distance is not expected to be significant.

Fig.~2a shows the relative perturbation, $\Delta\theta/\bar{\theta}$,
resulting from a 75~km- wide moving window.  We have checked that the
result is not affected by the choice of window size, by varying the
size from 55~km to 85~km.  Fig.~2b shows $N = [g\,{\rm
  d}(\ln\bar{\theta})/{\rm d}z]^{1/2}$ resulting from the
decomposition; here $g = g(z)$ is the gravity.  In the figure note
that $N$ is much smaller in the region below $\sim$20~km than that
above (the stratosphere)---in agreement with \citet{Magalhaes2002},
whose observations give $N \approx 6 \times 10^{-3}$~s$^{-1}$ in the
lower region.  This represents a possible ducting region, a source for
the waves in the probe data.

Although $w$ information was not collected by the probe, $m(z)$ can
still be obtained from $\Delta\theta$ through the polarization
relation \citep[e.g.,][]{Watkins2010}.  For this we generate a series
of Lomb-Scargle periodograms \citep{Scargle1982} of the $\Delta\theta$
data, which is non-uniformly spaced in $z$.  The non-uniform spacing
renders analysis by standard Fourier or wavelet transforms unsuitable.
One periodogram at each $z$ is generated using a smoothing window. We
have also verified in this procedure that the obtained result varies
little between different sized smoothing windows.  All the
periodograms are subsequently combined to produce a two-dimensional
map of the wave spectral energy density as a function of wavenumber
and altitude, ${\cal E} = {\cal E}(m^\ast,z)$, shown in Fig.~3; here
$m^\ast$ is the vertical wavenumber~$m$ prior to an adjustment for
wave propagation geometry.  Before discussing this adjustment, we
discuss three features which are already apparent in the ${\cal E}$
map.

\section{Results}

First, gravity waves identified previously \citep{Young2005} are
recovered (W1 and W2 in the map).  This agreement gives confidence in
our procedure.  In addition to those waves, we identify in our
analysis new gravity waves throughout the analyzed domain.  In
particular, note the high energy density (dark red) regions near the
50~km, 300~km and 400~km altitudes all with $m^\ast \approx 0.25$.
These constitute new waves.

Second, we identify a region consistent with a turbopause $\sim$50~km
thick, centered at approximately 400~km altitude.  This is the region
where the width of the sub-spectrum containing significant energy
increases markedly.  In this region, molecular diffusion becomes
comparable to eddy diffusion and gravity waves---growing in amplitude
as they propagate upward---break, transferring energy into the higher
wavenumbers.  Above this region, energy is lower across the entire
spectrum and the spectrum itself is much steeper (i.e., narrower) than
that for the turbopause region.  Above the turbopause region, the
atmosphere becomes inhomogeneous, separating out into layers of
different molecular species.  Past studies have placed Jupiter's
turbopause at the $\sim\!  5~\mu$bar \citep{Festou1981} and $\sim\!
0.5~\mu$bar \citep{Yelle1996} levels, based on observations and
modeling.  Our result supports the latter location.

Third, in the lower part of the analyzed domain there appears to be a
ducting region, a region with a sharp jump in $N$.  As already noted,
such a region can serve as a source of gravity waves.  Horizontally
propagating gravity waves in this region have previously been observed
(see, e.g., Fig.~1), which have been suggested as waves trapped in a
``leaky'' duct. The ducted wave travels horizontally by undergoing
internal reflections at the boundaries; however, part of the wave
escapes the duct to propagate vertically.  Because the wavenumber with
maximum energy in the ${\cal E}$ map can be traced down to the ducting
region, it is likely that the waves have come from there.

Note that above the ducting region the number of local peaks in the
spectrum generally reduces with altitude (the centroid of the spectrum
at each height is shifted to lower wavenumber).  Also, the magnitude
of the peak energy in the low wavenumber (white line in Fig.~3) is
high at first, then decreases, and then increases again along the
white line, until the topside of the turbopause region at $\sim\!
425$~km above the 1~bar level.  This is indicative of wave saturation
or encounters with a critical layer, where $I = 0$ locally, for high
number wavenumbers as they propagate upward.  The second or third
multi-peaked altitude regions, between $\sim\! 200$~km and $\sim\!
300$~km above the 1~bar level, may be due to wave breaking and
secondary wave generation from breaking layers.

As alluded to earlier, $m^\ast$ differs from the required wavenumber
$m$.  The latter is the wavenumber that would be observed by a probe
travelling in a vertical direction.  However,
throughout most of its entry phase, the probe had a shallow angle of
attack ($\sim\! 7^{\circ}$ below the horizontal), which changes the
wavenumber observed by the probe during its passage in the upper part
of the analyzed domain.  We use the geometry of the probe's path to 
obtain the true vertical wavenumber,
\begin{linenomath*}
\begin{equation}
  m\ =\ 
  \left(\frac{\tan\gamma\,\tan\beta}{1-\tan\gamma\,\tan\beta}\right)\,
  m^\ast,
\end{equation}
\end{linenomath*}
where $\gamma$ is the angle of attack and $\beta$ is the angle the
wavevector makes with the horizontal.  We make no correction for the
relative motion of the probe with respect to the wave since the
probe's velocity is supersonic (indeed, hypersonic, with up to Mach
51) for much of the entry phase.  Gravity wave phase speeds are
subsonic.

Although $\gamma = \gamma(z)$ is known from the probe's trajectory
\citep{Seiff1998}, $\beta$ is not.  To estimate this quantity, we
assume that the probe's trajectory is vertical in the period just
before parachute deployment.  This is not far from the actual
situation since the probe's angle of attack was $83^{\circ}$ just
before the parachute was deployed, near the lower boundary of our
domain.  In this region, $m^\ast \approx m$; and, since
\begin{linenomath*}
\begin{equation}
  \beta = \arccos\left( \frac{k}{\sqrt{k^2 + m^2}}\right)\, ,
\end{equation}
\end{linenomath*}
we have $\beta\approx 85^\circ$.  Also, in this region $m \approx
0.25$~km$^{-1}$ (giving vertical wavelength of $\approx\! 25$~km), in
good agreement with previously recovered values
\citep{Magalhaes2002,Arregi2009}.  Now, all the parameters required to
recover $I$ have been obtained, and a vertical profile for $I$ can be
estimated.

Finally, to obtain $u_0$ from $I$ a value for the horizontal phase
speed $c$ is required.  This is not well known.  However, equatorial
gravity waves with a phase speed of about 100~m~s$^{-1}$ greater than
the winds have been observed in images of Jupiter's clouds returned by
New Horizons \citep{Reuter2007}.  This gives $c \approx
180$~m~s$^{-1}$.  The Doppler Wind Experiment (DWE) on the probe has
measured the zonal winds to be approximately this speed in the deeper
atmosphere \citep{Atkinson1998}, suggesting that the waves could have
been driven by convective overshoot below the clouds.  This value also
agrees well with the 140--195~m~s$^{-1}$ range reported for the
similar region, at the probe entry site \citep{Magalhaes2002}.
Somewhat smaller values (approximately 70--145~m~s$^{-1}$) have been
obtained for waves at different longitudes \citep{Arregi2009}.  Such
variation in $c$ does not change the shape of the $u_0$
profile we recover, but the magnitude would be reduced.  Variations of
$c$ with altitude would increase the uncertainty as well, but they are
expected to be small and not fundamentally change the profile.  This
gives the profile for $u_0(z)$ shown in Fig.~4a.  Here we have used $c
= 180$~m~s$^{-1}$.  The standard deviation shown indicates the
variation in the profile given by the various smoothing window sizes
considered in constructing ${\cal E}$.

\section{Discussion}

The DWE reported flow speeds that increase with depth reaching
$170$~m~s$^{-1}$ at the $5$~bar level and then remaining high at lower
levels \citep{Atkinson1998}.  The upper part of the DWE profile is
shown in Fig.~4a.  The wind speed at the bottom of our profile
agrees well with that at the top of the DWE profile.  Our profile
shows increasing zonal wind speed with altitude (up to $\sim\! 100$~km
above the 1~bar level).  This implies that the wind speed is a minimum
near the cloud-top level. This is similar to what has been observed
from studies using the thermal wind equation
\citep{Flasar2004,SimonMiller2006}.  However, in our profile this high
speed wind is not a jet, as the zonal wind speed does not diminish
appreciably with altitude near this level.  In fact, the speed is
roughly constant throughout the stratosphere, starting from this
level.  There are some fluctuations of the order of 20~m~s$^{-1}$ in
the wind speed in the thermosphere.  The Richardson number, $Ri =
N^2\,({\rm d}u_0/{\rm d}z)^{-2}$, is greater than $1/4$ for the entire
profile, which indicates the flow is stable with respect to
Kelvin-Helmholtz instability.

Temporal variations in the temperature profile of Jupiter's equatorial
stratosphere have been observed to have a period of $4$ to $5$ years
and are therefore known as the quasi-quadrennial oscillation (QQO)
\citep{Leovy1991,Friedson1999}.  These oscillations have been linked
to variations in the zonal wind observed via cloud-tracking on Jupiter
with a period of $\sim$4.4~years \citep{SimonMiller2007}.  Further, a
jet located just north of the equator, derived using the thermal wind
equation applied to data gathered in December 2000 and January 2001,
has been linked to the QQO \citep{Flasar2004}.  A jet of similar
magnitude, just south of the equator, has been observed in
observations gathered in 1979 \citep{SimonMiller2006} (which fits well
with a period of 4.4~years), suggesting this jet may be oscillating
about the equator.  No such jet is visible in our profile.  However,
the probe entered Jupiter's atmosphere 5 years before the northern jet
was observed.  Thus, our profile is consistent with such an
oscillation as the jet would be located south of the equator, away
from the probe entry site, at the time the probe entered.

Gravity wave creation and dissipation is associated with heating and
cooling of the atmosphere.  Ignoring molecular viscosity, which is
only important above the turbopause, the heating rate is proportional
to the vertical gradient of energy flux---specifically,
\begin{linenomath*}
\begin{equation}
\frac{\partial T}{\partial t} = - \frac{1}{\rho c_p} \frac{\partial F_z}
{\partial z}
\end{equation}
\end{linenomath*}
We estimate the energy flux, $F_z=\rho \overline{\phi w}$, from
$\Delta \theta$ using the polarization relations to derive the
perturbations in the geopotential $\phi$ and the vertical velocity
$w$.  The flux comes from the zonal average of the product of these
quantities.  Since the product is not available to us we average over
a single vertical wavelength instead.  The heating rate profile thus
obtained is shown in Fig~4b.  It can be seen that heating is small
through the stratosphere.  Indeed, on average it is very close to
zero.  It is only in the thermosphere that the average heating rate
deviates from zero with a magnitude of $\sim\!$ 0.5~K per Jupiter
rotation.  The flux responsible for heating in this region is small
($F_z < 10^{-3}$~W~m$^{-2}$), compared to radiation fluxes measured in
the troposphere ($3 < F_z < 16$~W~m$^{-2}$) \citep{Sromovsky1998}.
Nevertheless, the peak heating rate magnitude is about 50 times
larger, partly because the lower thermsopshere is much less dense than
the troposphere.

The circulation of Jupiter's stratosphere is important for
understanding the planet's circulation as a whole.  The location of
the turbopause is essential for understanding the coupling between
Jupiter's upper atmosphere and the circulation in the lower
atmosphere.  The Juno mission will explore the troposphere of Jupiter
to depths of $100$~bar or more.  This will provide better insight to
the source and behaviour of gravity waves in the troposphere, possibly
allowing better limits to be derived for the wave-vector angle $\beta$
in our analysis and better understanding of the mechanisms that
generate the gravity waves.  The planned Jupiter Icy Moon Explorer
(JUICE) mission will directly study gravity wave activity and zonal
winds in the stratosphere of Jupiter, extending the result present
here.

%%% End of body of article:

%%%%%%%%%%%%%%%%%%%%%%%%%%%%%%%%
%% Optional Appendix goes here
%
% \appendix resets counters and redefines section heads
% but doesn't print anything.
% After typing  \appendix
%
% \section{Here Is Appendix Title}
% will show
% Appendix A: Here Is Appendix Title
%
%%%%%%%%%%%%%%%%%%%%%%%%%%%%%%%%%%%%%%%%%%%%%%%%%%%%%%%%%%%%%%%%
%
% Optional Glossary or Notation section, goes here
%
%%%%%%%%%%%%%%
% Glossary is only allowed in Reviews of Geophysics
% \section*{Glossary}
% \paragraph{Term}
% Term Definition here
%
%%%%%%%%%%%%%%
% Notation -- End each entry with a period.
% \begin{notation}
% Term & definition.\\
% Second term & second definition.\\
% \end{notation}
%%%%%%%%%%%%%%%%%%%%%%%%%%%%%%%%%%%%%%%%%%%%%%%%%%%%%%%%%%%%%%%%
%
%  ACKNOWLEDGMENTS

\begin{acknowledgments}
  C.W. is supported by the Science and Technology Facilities Council
  (STFC), and J.Y-K.C. is supported by the STFC PP/E001858/1 grant.
  The authors acknowledge H.~Thrastarson, I. Polichtchouk, and
  C. Agnor for useful discussions.  We thank the reviewers for
  helpful suggestions.
\end{acknowledgments}

\end{article}

%% Enter Figures and Tables here:

% When submitting articles through the GEMS system:
% COMMENT OUT ANY COMMANDS THAT INCLUDE GRAPHICS.
%
% FOR FIGURES, DO NOT USE \psfrag or \subfigure commands.
%
% Figure captions go below the figure.
% Table titles go above tables; all other caption information
%  should be placed in footnotes below the table.
\newpage
\begin{figure}%fig1
  \begin{center}
 \noindent\includegraphics[width=20pc]{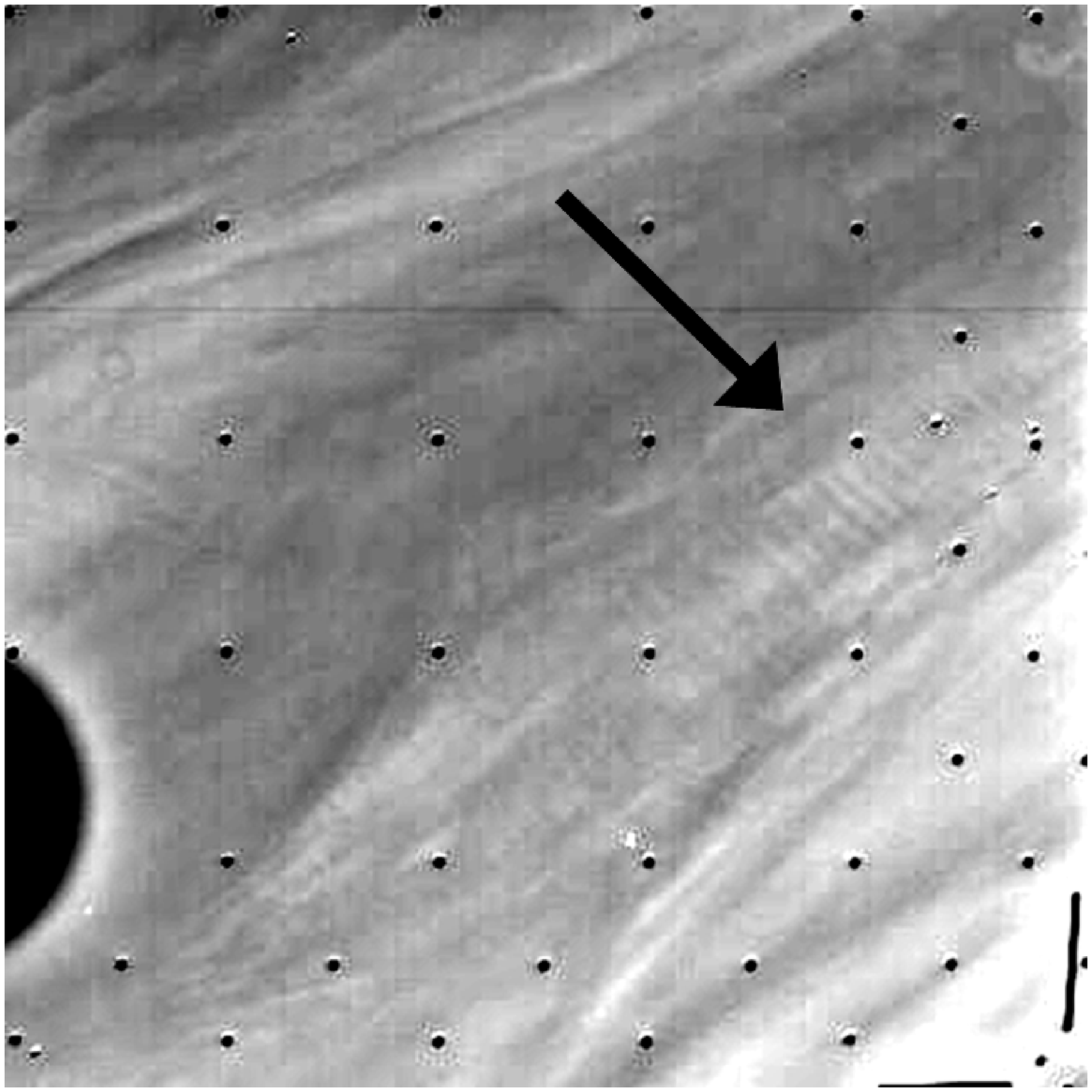}
  \end{center}
  \caption{Gravity wave in Jupiter's atmosphere (Voyager image 16316.34).
    This wave is propagating horizontally in the troposphere.  Many such 
    waves are captured in the Voyager images \citep{Flasar1986}.  Image 
    contrast has been enhanced to improve the visibility of the wave.  
    The wave's location is indicated by the arrow. Image courtesy 
    NASA/JPL-Caltech.}
\end{figure}

\begin{figure}%fig2
  \begin{center}
\noindent\includegraphics[width=20pc]{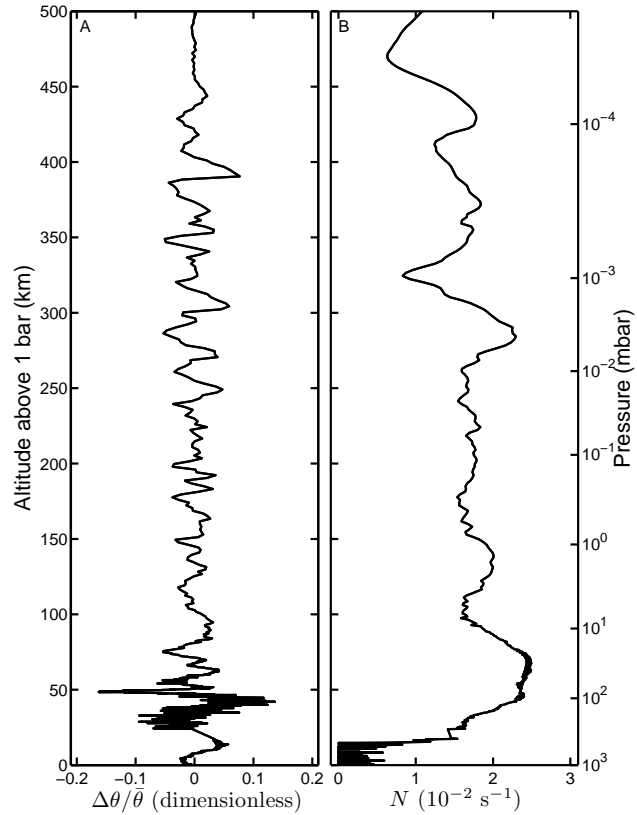}
  \end{center}
  \caption{Relative potential temperture perturbation and
    Brunt--V\"ais\"al\"a (buoyancy) frequency profiles for Jupiter's
    atmosphere. All altitudes are relative to the 1~bar pressure
    level. The profiles are based on data gathered from $t =
    -173.055$~s to $t = 218.170$~s; $t = 0$~s is the point of
    parachute deployment. Our analysis used the complete set of
    acceleration data collected from both accelerometers with the
    exception of one outlier at $t = -157.742$~s. The acceleration
    data collected by the probe includes a small spurious
    oscillation\cite{Seiff1998} which we did not smooth out as its
    effects on the calculation was negligible in the atmospheric
    region we analyzed. The bottom 25~km of the profile is from
    direct measurement of the temperature with the first 15~s of
    direct temperature measurements removed as these were anomalously
    high.  \textbf{a,} The vertical profile of potential temperature 
    perturbations, $\Delta \theta$, scaled by the background value, 
    $\bar{\theta}$. The perturbations show wavelike oscillations
    throughout the stratosphere and lower thermosphere. Note that the
    short wavelength oscillations in the layer between 25~km and 50~km
    are due to accelerations caused by buffeting of the probe as its
    velocity became subsonic. \textbf{b,} The Brunt--V\"ais\"al\"a
    frequency is the maximum frequency that a gravity wave can have
    and shows the stability of the atmosphere against convective
    stability.}
\end{figure}

\begin{figure}%fig3
  \begin{center}
 \noindent\includegraphics[width=20pc]{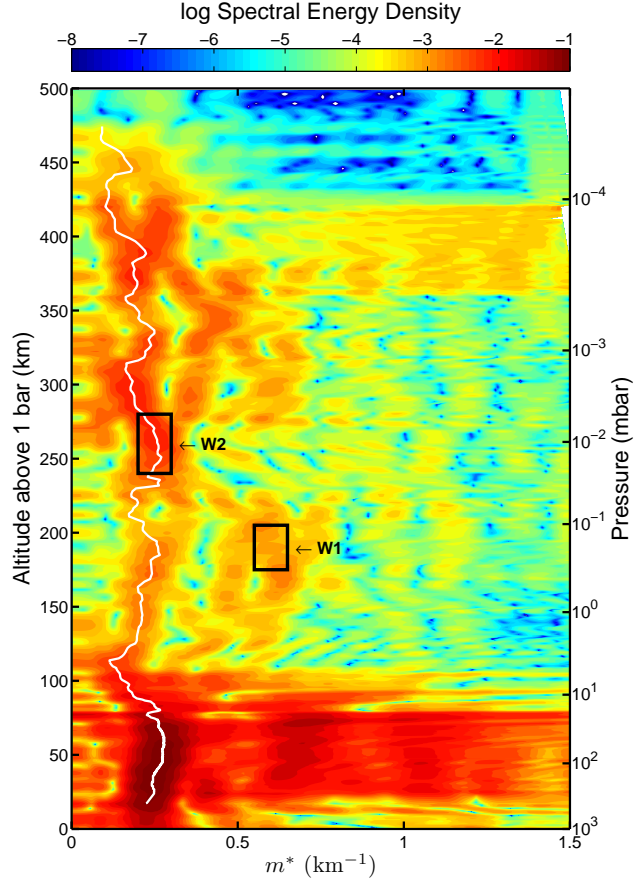}
  \end{center}
  \caption{Moving Lomb-Scargle periodogram of the potential
    temperature perturbation from Fig.~2a. Regions of high spectral
    energy are in red while low energy is in blue. The variation of
    the observed vertical wavenumber $m^\ast$ is shown (white
    line). The line is produced by identifying local energy maxima
    and constructing a line joining them, avoiding local minima, 
    starting at the level of the duct in the troposphere. Note that 
    there are other waves within the periodogram we do not consider 
    in our analysis. One such wave,     the region of high energy at
    $m^\ast\approx 0.61$~km$^{-1}$ between 180~km and 210~km altitude
    (labelled W1) has been previously identified as a saturating gravity
    wave\cite{Young2005}.  Another region (labelled W2) has also been
    previously identified\cite{Young2005}.  The region at an altitude
    of around 400~km shows a broadening of the range of wavenumbers
    with increased spectral energy. This is indicative of the
    turbopause, the region where waves break and the atmosphere begins
    to become heterogeneous.}
\end{figure}

\begin{figure}%fig4
\begin{center}
\noindent\includegraphics[width=20pc]{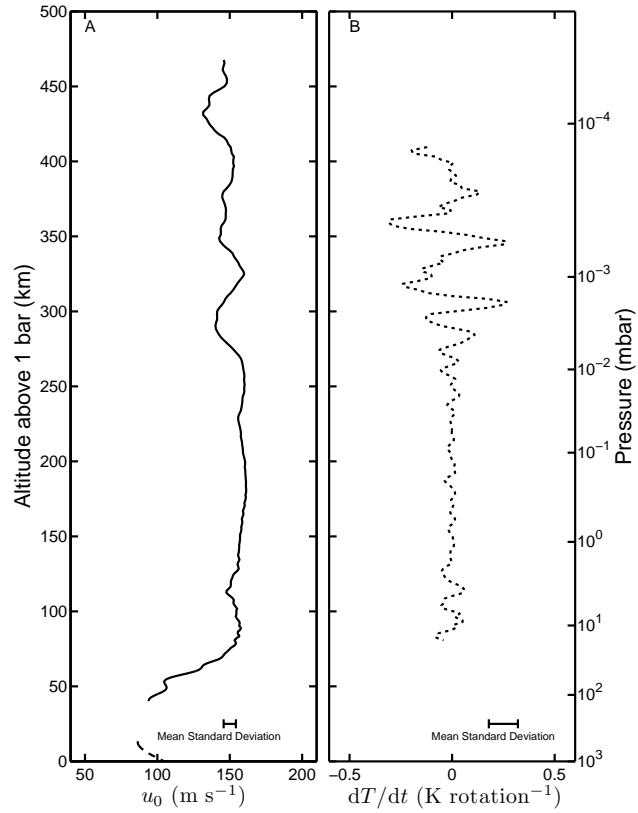}
\end{center}
\caption{The vertical profiles of the background zonal wind speed
  $u_0$ and heating rate, ${\rm d}T/{\rm d}t)$.  \textbf{a,} The zonal
  wind is shown (solid) with the average standard deviation of the
  variation across smoothing and periodogram windows indicated. The
  zonal speed profile found by the Doppler Wind Experiment is shown at
  the bottom (dashed) for comparison.  \textbf{b,} The vertical profile
  of the heating rate is shown (dotted), as temperature change per
  Jovian rotation (9.925~h), with the average standard deviation of
  the variation across smoothing and periodogram windows indicated. }
 \end{figure}

% DRAFT figure/table, including eps graphics
%
% \begin{figure}
% \noindent\includegraphics[width=20pc]{samplefigure.eps}
% \caption{Caption text here}
% \end{figure}
% \end{document}
%
% \begin{table}
% \caption{}
% \end{table}
%
% ---------------
% TWO-COLUMN figure/table
%
% \begin{figure*}
% \noindent\includegraphics[width=39pc]{samplefigure.eps}
% \caption{Caption text here}
% \end{figure*}
%
% \begin{table*}
% \caption{Caption text here}
% \end{table*}
%
% ---------------
% EXAMPLE TABLE
%
%\begin{table}
%\caption{Time of the Transition Between Phase 1 and Phase 2\tablenotemark{a}}
%\centering
%\begin{tabular}{l c}
%\hline
% Run  & Time (min)  \\
%\hline
%  $l1$  & 260   \\
%  $l2$  & 300   \\
%  $l3$  & 340   \\
%  $h1$  & 270   \\
%  $h2$  & 250   \\
%  $h3$  & 380   \\
%  $r1$  & 370   \\
%  $r2$  & 390   \\
%\hline
%\end{tabular}
%\tablenotetext{a}{Footnote text here.}
%\end{table}

% See below for how to make landscape/sideways figures or tables.

\end{document}